\documentclass[aps,pra,10pt,showpacs,twocolumn,superscriptaddress,longbibliography]{revtex4-1}
\usepackage{graphicx}
\usepackage[usenames]{color}
\usepackage{amssymb,amsmath,amsfonts}
\usepackage{siunitx}
\usepackage{xcolor}
\usepackage{setspace} 
\usepackage{float} 
\usepackage{tabularx}
\usepackage[linkcolor=blue,citecolor=blue,colorlinks=true,urlcolor=blue]{hyperref}
\newcommand{\eq}[1]{Eq.~(\ref{#1})}
\newcommand{\fig}[1]{Fig.~\ref{#1}}
\newcommand{\be}[1]{\begin{equation}\label{#1}}
\newcommand{\ee}{\end{equation}}

\usepackage{lipsum}

\begin{document}

\title{Momentum scalar triple  product as a measure of  chirality in  electron ionization dynamics of strongly-driven atoms}

\author{G. P. Katsoulis}
\affiliation{Department of Physics and Astronomy, University College London, Gower Street, London WC1E 6BT, United Kingdom}

\author{Z. Dube}
\affiliation{Joint Attosecond Science Lab of the National Research Council and the University of Ottawa, Ottawa, Ontario K1A 0R6, Canada}

\author{P. Corkum}
\affiliation{Joint Attosecond Science Lab of the National Research Council and the University of Ottawa, Ottawa, Ontario K1A 0R6, Canada}

\author{A. Staudte}
\affiliation{Joint Attosecond Science Lab of the National Research Council and the University of Ottawa, Ottawa, Ontario K1A 0R6, Canada}

\author{A. Emmanouilidou}
\affiliation{Department of Physics and Astronomy, University College London, Gower Street, London WC1E 6BT, United Kingdom}

\begin{abstract}
We formulate a transparent measure that quantifies chirality in single electron ionization triggered in atoms, which are achiral systems. We do so in the context of Ar driven  by a new type of optical fields that consists of 
two non-collinear laser beams giving rise to chirality that varies in space across the focus of the beams. Our computations  account for realistic experimental conditions. To define this measure of chirality, we first find  the sign of the electron final momentum scalar triple product  $\mathrm{{\bf{p}}_{k}\cdot ({\bf{p}}_{i}\times {\bf p}_{j})}$ and multiply it with the probability for an electron to ionize with certain values for  both $\mathrm{p_{k}}$ and  $\mathrm{p_{i}p_{j}}$. Then, we integrate this product over all  values of $\mathrm{p_{k}}$ and  $\mathrm{p_{i}p_{j}}$.  We show this to be 
 a robust measure of chirality in electron ionization triggered by globally chiral electric fields.
\end{abstract}

\date{\today}

\maketitle


Ultrafast phenomena in chiral molecules triggered by intense, infrared laser pulses are at the forefront of laser-matter interactions \cite{Chiral4,Beaulieu2017Science,Beaulieu2018NatPhys,Comby2018NatComm,Chiral5}. While ultrafast chiral processes can be studied using high harmonic generation (HHG) \cite{Chiral4,Chiral6,Baykusheva2018PRX,Neufeld2018PRL,Neufeld2019PRX,Heinrich2021NatComm}, the underlying recollision mechanism entails that a stronger chiral response comes at the expense of a greatly suppressed high harmonic signal \cite{Chiral4}. Hence, photoelectron spectroscopy is a promising route to a robust signal from molecules driven by intense chiral fields \cite{Lux2012AngChem,StefanLehmann2013JCP,Chiral2,Beaulieu2017Science,Beaulieu2018NatPhys,Chiral5}. However, the sensitivity of chiral photoelectron spectroscopy also struggles with the fact that laser wavelengths are several orders of magnitude larger than the molecular dimensions, i.e., the chiralities of the optical field and the molecule are incommensurate.
 
\begin{figure}[t]
\centering
\includegraphics[width=1\linewidth]{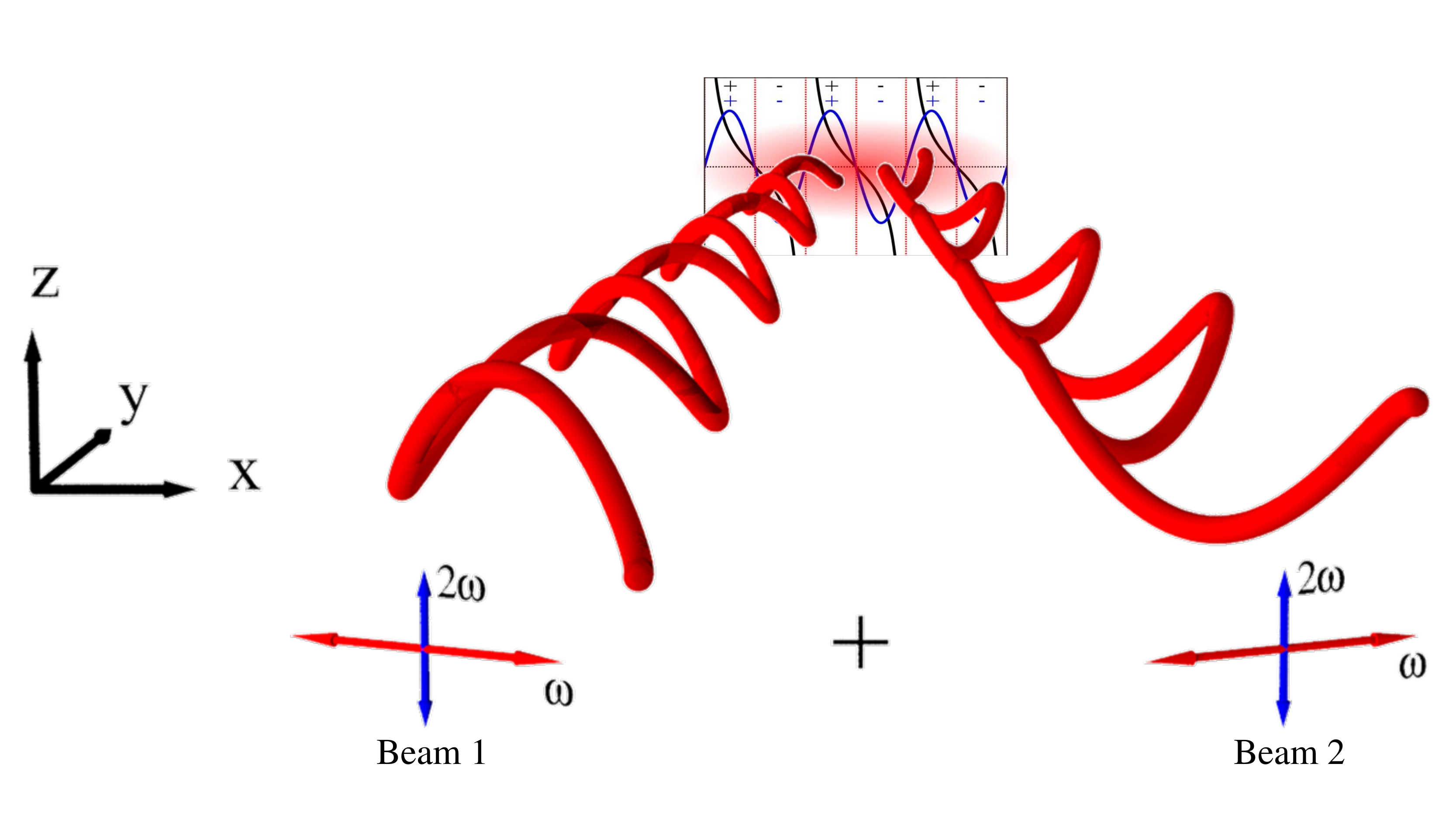}
\caption{Schematic plot of two non-collinear laser beams, each consisting of $\omega$-$2\omega$ orthogonally polarized linear fields; $\omega$ is polarized on the x-y plane and 2$\omega$  along the z-axis. Both  beams propagate towards the atom in the focus region (red-shaded ellipse). The resultant electric field has chirality  (square inset) that changes  along the x-axis in the focus region.} \label{fig:fig1}
\end{figure}

Recently, Ayuso et al. proposed a new type of optical field which is chiral on the atomic scale \cite{Chiral6} and thereby holds the potential for unprecedented chiral sensitivity. The chiral field is synthesized by  combining two orthogonally polarized two-color laser fields in a non-collinear geometry as illustrated in  Fig. \ref{fig:fig1}. The non-collinear geometry creates an intensity and ellipticity grating, and thereby causes the chirality of the laser field to spatially vary across the focus. Thus, it is a fundamental challenge for experiments to decipher the signatures of chirality in the photoelectron spectra from these new laser fields.


 Here, we provide a simple prescription on how to analyze experimental photoelectron spectra produced from any type of chiral light. To this end, we perform semi-classical simulations of strong-field ionization, taking into account the focal volume distribution of the degree of light chirality. To develop an understanding of chiral electron ionization we model atomic photoionization, since ground-state atoms have spherical symmetry and are intrinsically achiral systems. Thus, the chiral response of the escaping electron that is imprinted on the ionization spectra in our model arises solely from the dynamics triggered by the electric field of the laser.

In order to analyze the resulting photoelectron spectra we identify a transparent measure that quantifies chirality in electron  ionization dynamics ensuing from an atom strongly-driven by a chiral electric field. We construct this measure by finding  the probability 
P$(\mathrm{p_{k}, p_{i}p_{j}})$ for an electron to ionize with certain values for both $\mathrm{p_{k}}$ and $\mathrm{p_{i}p_{j}}$, with p$_{\mathrm{i}}$,p$_{\mathrm{j}}$,p$_\mathrm{k}$ being the  components of the final electron momentum. Next, we multiply this probability by the sign of the momentum scalar triple product $\mathrm{{\bf{p}}_{k}\cdot ({\bf{p}}_{i}\times {\bf p}_{j})}$. Integrating over the  whole range of values of $\mathrm{p_{k}}$ and $\mathrm{p_{i}p_{j}}$, we obtain a measure of chirality 
\begin{align}
\begin{split}
\mathcal{X}\mathrm{(p_{k},p_{i}p_{j})} &=\\
& \mathrm{\iint sign(\mathrm{{\bf{p}}_{k}\cdot ({\bf{p}}_{i}\times {\bf p}_{j})}) P(\mathrm{p_{k}, p_{i}p_{j}}) dp_{i}p_{j}dp_{k} }.\label{eq:main}
\end{split}
\end{align}
We show that  $\mathcal{X}$ has an opposite sign for synthetic pulses with opposite chirality, while it is zero for  achiral synthetic pulses. This measure of handedness of electron ionization dynamics is a general one and  accounts for chiral electron motion triggered by any chiral light.

We demonstrate that  $\mathcal{X}$ is a measure of handedness of electron ionization dynamics ensuing from atoms, in the context of Ar  driven by  two non-collinear laser beams, see \fig{fig:fig1}. Beams 1, 2   propagate on the x-y plane with wavevectors $\mathbf{k}_1$, $\mathbf{k}_2$  forming an angle $\alpha$ with the y axis
\begin{align}
\begin{split}
\mathbf{k}_1 &= \mathrm{k}\sin(\alpha)\hat{\mathbf{x}} +  \mathrm{k}\cos(\alpha)\hat{\mathbf{y}} \\
\mathbf{k}_2 &= -\mathrm{k}\sin(\alpha)\hat{\mathbf{x}} +  \mathrm{k}\cos(\alpha)\hat{\mathbf{y}},
\end{split}
\end{align}
where $k = \mathrm{2 \pi} / \lambda $.
The electric field of each beam consists of two orthogonally polarized  $\omega$ and $2\omega$ laser fields. The $\omega$ field is polarized along the x-y plane and the $2\omega$ field is polarized along the z-axis. Also, we take
the 2$\omega$ field to have small intensity compared to the $\omega$ field.

The resultant electric field is given by \cite{Chiral6}
\begin{equation}\label{eq:Electric_field}
\mathbf{E}(\mathbf{r},\mathrm{t})=2\mathrm{E}_0 \exp \left[ - \left( \frac{t}{\tau} \right)^2\right]\left( \mathrm{E_x} \hat{\mathbf{x}} + \mathrm{E_y} \hat{\mathbf{y}}  + \mathrm{E_z} \hat{\mathbf{x}}   \right),
\end{equation}
where
{\allowdisplaybreaks
\begin{align}
\begin{split}
\mathrm{E_{x/y}}&=\exp \left[- \left( \frac{ \rho }{\mathrm{w_{0}}} \right)^2 \right]\mathrm{f_{x/y}(x)} \cos\left[\mathrm{g(y,t)}\right]\\
\mathrm{E_z}&=\exp \left[- \left( \frac{ 2\rho }{\mathrm{w_{0}}} \right)^2 \right]\mathrm{f_z(x)} \cos\left[\mathrm{h(y,t)}\right],
\end{split}
\end{align}}
and  
{\allowdisplaybreaks
\begin{align}
\begin{split}
\mathrm{f_x(x)} &= \cos(\alpha)\cos \left[\mathrm{k}\sin(\alpha)x+\frac{\phi_2^{\omega}-\phi_1^{\omega}}{2} \right]\\
\mathrm{f_y(x)} &= \sin(\alpha)\sin \left[\mathrm{k}\sin(\alpha)x+\frac{\phi_2^{\omega}-\phi_1^{\omega}}{2} \right]\\
\mathrm{f_z(x)} &=\mathrm{r_0}\cos \left[2\mathrm{k}\sin(\alpha)x + (\phi_2^{2\omega}-\phi_1^{2\omega})\right]\\
\mathrm{g(y,t)}&= \mathrm{k}\cos(\alpha) \mathrm{y} -\omega \mathrm{t} - \frac{\phi_2^{\omega}+\phi_1^{\omega}}{2}\\
\mathrm{h(y,t)}&=2\mathrm{k}\cos(\alpha) \mathrm{y} -2\omega \mathrm{t} - (\phi_2^{2\omega}+\phi_1^{2\omega}).
\end{split}
\end{align}}
We note that $\tau = 25$ fs and $\tau\sqrt{2\ln(2)}$ is the full width at half maximum of the pulse duration in intensity, while  $\mathrm{E_0}$ is the field strength  corresponding to intensity $\mathrm{5 \times 10^{13} \; W/cm^2}.$ 
Also, $\rho$   is the radial distance to the propagation axis of each laser beam. Since  $\alpha$ is small,  $5^{\circ}$, it follows that $\rho   \approx \sqrt{\mathrm{x}^2 + \mathrm{z}^2}$. Moreover,   $\mathrm{w_{0} = 8.49 \; \mu m}$ is the beam waist of the $\omega$ laser field,  and  $\mathrm{r^2_0}$ is the intensity ratio of 1/100 of the 2$\omega$ versus the $\omega$ field. Finally,  the wavelength $\lambda$   of the $\mathrm{\omega}$ field is taken equal to 800 nm.

We treat single electron ionization of driven Ar by employing a three-dimensional (3D) semi-classical model.  The only approximation is  the initial state. One electron  tunnel-ionizes  through the field-lowered Coulomb-barrier at time t$_{0}$.  To compute the  tunnel-ionization rate,  we employ the quantum mechanical Ammosov-Delone-Krainov (ADK) formula  \cite{A1,A2}.     We  use parabolic coordinates to obtain the exit point of the tunneling electron   along the laser-field direction \cite{parabolic1}.  We set the electron momentum  along the laser field  equal to zero, while  we obtain the transverse momentum  by a Gaussian distribution  \cite{A1,A2}.  The  microcanonical distribution is employed to describe the initial state of the initially bound electron \cite{Abrimes}. 

 We select the tunnel-ionization time, t$_{0}$, randomly  in the time interval  [-2$\tau$,2$\tau$]. Next, we specify at time $\mathrm{t_{0}}$ the initial conditions for both electrons. Using the three-body Hamiltonian of the two electrons with the nucleus kept fixed, we propagate classically in time  the position and momentum of each electron. All Coulomb forces and the interaction of each electron with the electric field in \eq{eq:Electric_field} are fully accounted for with no approximation. To account for the Coulomb singularity, we employ  regularized coordinates \cite{KS}. Here, we use atomic units.
  
  Previous successes of this model include identifying the mechanism underlying the fingerlike structure in the correlated electron momenta for He driven by 800 nm laser fields \cite{Emm1}, see also \cite{Taylor1,exp1,exp2}. Moreover, we  investigated the direct versus the delayed pathway of non-sequential double ionization for He driven by a 400 nm, long duration laser pulse and achieved excellent agreement with fully ab-initio quantum mechanical calculations \cite{Emm2}. Also, for low intensities, we have  identified a novel mechanism of double ionization, namely, slingshot non-sequential double ionization \cite{Emm4}. In addition, for several observables of non-sequential double ionization, our results have good agreement with experimental results  for Ar when driven by near-single-cycle laser pulses at 800 nm \cite{Emm3}.

 It was previously shown \cite{Chiral6} that the resultant electric field  is globally chiral if the relative phases  of the $\omega$ and $2\omega$ laser fields in  beams 1 and 2,  i.e.  $\phi_{1}^{2\omega}-\phi_1^{\omega}$ and $\phi_{2}^{2\omega}-\phi_2^{\omega}$, satisfy the following condition   
\begin{equation}\label{eq:chiral}
\mathrm{\left( \phi^{2\omega}_{1} - \phi^{\omega}_{1}\right) - \left( \phi^{2\omega}_{2} - \phi^{\omega}_{2}\right) = \dfrac{\pi}{2} + n \pi, \text{with n} \in} \mathbb{Z}, 
\end{equation}
while the resultant electric field is globally achiral when
\begin{equation}\label{eq:achiral}
\mathrm{\left( \phi^{2\omega}_{1} - \phi^{\omega}_{1}\right) - \left( \phi^{2\omega}_{2} - \phi^{\omega}_{2}\right) = n \pi, \text{with n} \in} \mathbb{Z}.
\end{equation}

\begin{figure}[t]
\centering
\includegraphics[width=0.7\linewidth]{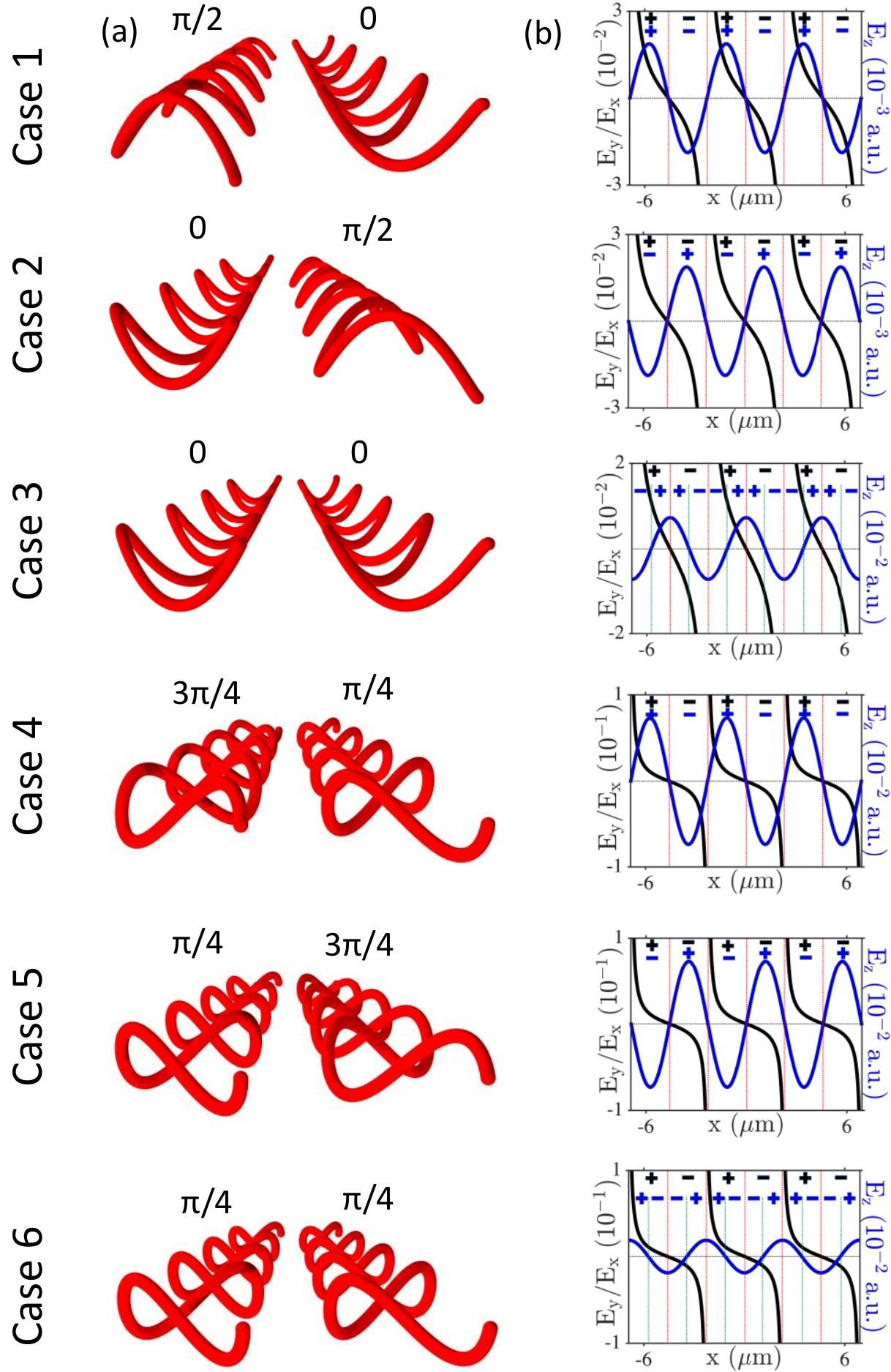}
\caption{(a) Schematic plots of six combinations of two non-collinear beams; (b) change of sign of  E$\mathrm{_y}$(x)/E$\mathrm{_x}$(x) (black) and E$\mathrm{_z}$(x) (blue) along the x-axis in the focus region, at $\text{y=z=0}$, $\text{t=T/50}$. Cases 1,2  correspond to globally chiral electric  fields with opposite handedness, the same holds for cases 4 and 5. Cases 3 and 6 correspond to globally achiral electric fields. For each case, above beams 1 and 2, we denote  $\phi_{1}^{2\omega}-\phi_1^{\omega}$ and $\phi_{2}^{2\omega}-\phi_2^{\omega}$, respectively.}
\label{fig:Fig2}
\end{figure}

To illustrate that $\mathcal{X}$ is a measure of chirality in electron ionization of driven atoms, we perform six independent studies.  Each study  corresponds to Ar being driven by one of six different resultant electric fields corresponding to six different synthetic pulses. For simplicity,  we refer to the resultant electric field of the synthetic pulse as electric field. Each of the six synthetic pulses (cases 1-6) corresponds to a different combination of $\phi_{1}^{2\omega}-\phi_1^{\omega}$ and $\phi_{2}^{2\omega}-\phi_2^{\omega}$ for  beams 1 and 2, respectively, see \fig{fig:Fig2}(a).
Using the conditions in \eq{eq:chiral} and \eq{eq:achiral}, we select four globally chiral electric fields, cases  1,2,4,5, and two globally achiral fields, cases 3, 6,   see \fig{fig:Fig2}. 
 In \fig{fig:Fig2}(b), we show that the  electric fields which are globally chiral maintain the same handedness along the x-axis in the focus region. That is,   E$\mathrm{_y}$(x)/E$\mathrm{_x}$(x) and E$\mathrm{_z}$(x) change sign at the same points in space x. As a result, electric fields 1 and 4  have the same chirality (+) in \fig{fig:Fig2}(b) and electric fields  2 and 5  have the same chirality (-) in \fig{fig:Fig2}(b). It follows that the pairs of  electric fields (1,2) and (4,5) have opposite chirality. Also, when  E$\mathrm{_y}$(x)/E$\mathrm{_x}$(x) and E$\mathrm{_z}$(x) change sign at different points in space x as defined by \eq{eq:achiral}, the chirality of the electric field flips sign along the x-axis in the focus region.  Hence, the  electric field  has no overall chirality. This is the case for the globally achiral  fields 3 and 6 shown in \fig{fig:Fig2}(b).

Next, we describe how we obtain the electron ionization spectra of Ar for each of the six synthetic pulses (cases 1-6). For simplicity,  for each case,   we set $\phi_2^{\omega} = 0$. Since only the differences  $\phi_{1}^{2\omega}-\phi_1^{\omega}$ and $\phi_{2}^{2\omega}-\phi_2^{\omega}$ are important, there is no loss of generality.  Moreover, for each of the six synthetic pulses, to simulate realistic experimental conditions, we select 101 equally spaced values of the phase $\mathrm{\phi_1^{\omega}}$ in the interval [0,2$\pi$). This allows us to  account for the nucleus being at different positions along the x-axis  in the focus region. Next, for each of  the 101 values of $\mathrm{\phi_1^{\omega}}$, we register the single ionisation events and obtain the electron ionization spectra. Then, we average over all $\mathrm{\phi_1^{\omega}}$ values and obtain the electron spectra $\mathrm{P^m}(\mathrm{p_{x}},\mathrm{p_{y}p_{z}})$, $\mathrm{P^m}(\mathrm{p_{y}},\mathrm{p_{z}p_{x}})$ and $\mathrm{P^m}(\mathrm{p_{z}},\mathrm{p_{x}p_{y}})$. We normalize each one of these spectra to one. The m index corresponds to the m electric  field, i.e. to case m and ranges from 1-6.  For each synthetic pulse 1-6, the electron ionization spectra  are obtained using  at least 10$^{7}$ singly ionizing trajectories. 

\begin{figure}[t]
\centering
\includegraphics[width=1.\linewidth]{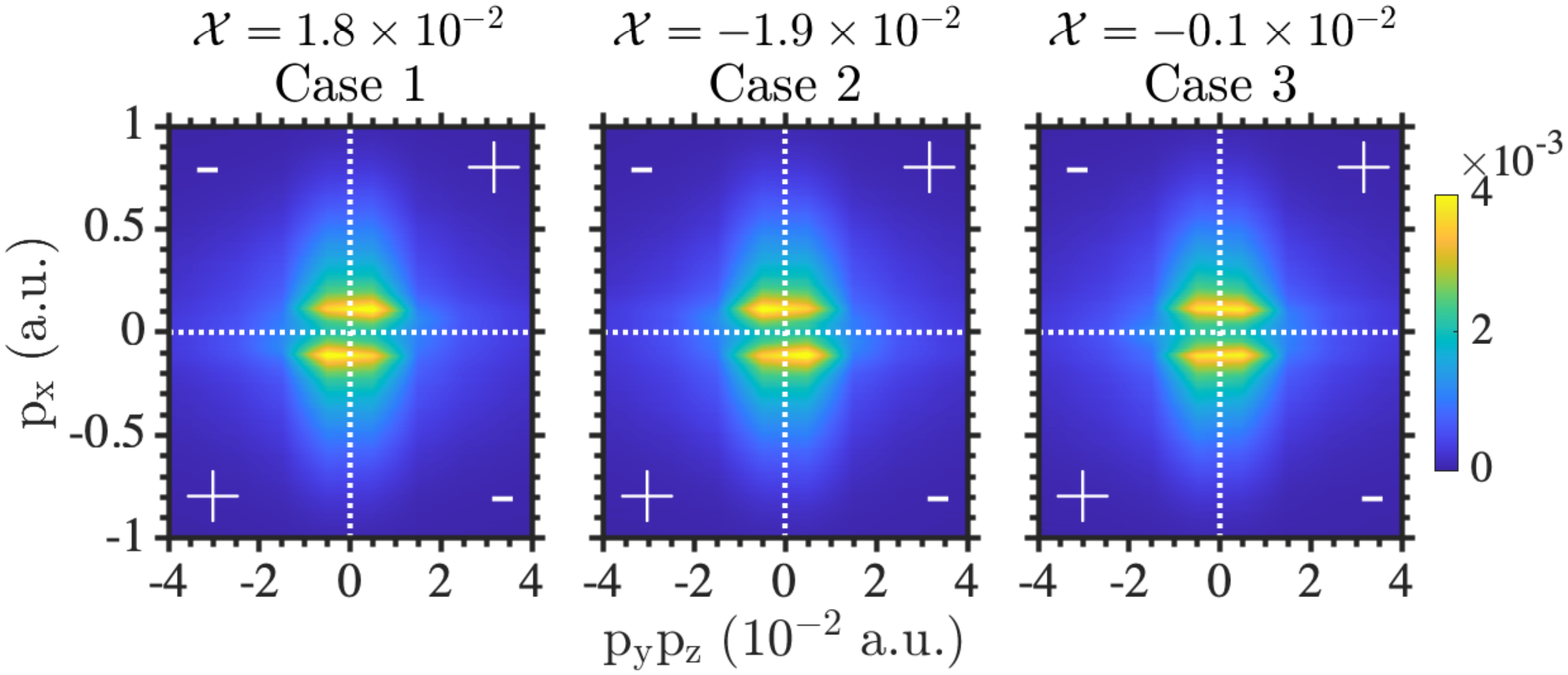}
\caption{Probability distribution $\mathrm{P^m}(\mathrm{p_{x}},\mathrm{p_{y}p_{z}})$ for the  electron to ionize with momenta  $\mathrm{p_x}$ and  $\mathrm{p_yp_z}$, with m=1,2,3 for electric fields 1,2,3, respectively. The sign in each quadrant corresponds to the sign of $\mathrm{{\bf{p}}_{x}\cdot ({\bf{p}}_{y}\times {\bf p}_z)}$ in this quadrant. }
\label{fig:fig3}
\end{figure}

In \fig{fig:fig3}, we plot the probability distribution $\mathrm{P^m}(\mathrm{p_{x}},\mathrm{p_{y}p_{z}})$ for an electron to singly ionize with momenta p$\mathrm{_{x}}$ and p$\mathrm{_{y}p_{z}}$ for the globally chiral 1, 2 and globally achiral 3 electric fields. In each quadrant we assign the sign resulting from the scalar triple product $\mathrm{{\bf{p}}_{x}\cdot ({\bf{p}}_{y}\times {\bf p}_z)}$. Next, using \eq{eq:main}, we compute  the measure of chirality in electron ionization $\mathcal{X}(\mathrm{p_{x}},\mathrm{p_{y}p_{z}})$. We find it to be equal to $1.8\times10^{-2}, -1.9\times10^{-2},-0.1\times10^{-2}$ for electric fields 1,2,3, respectively.  
 Indeed, a close inspection of \fig{fig:fig3} for case 1 reveals that  the probability distribution of the electron momenta p$\mathrm{_{x}}$ and p$\mathrm{_{y}p_{z}}$ has larger values  at the first and third quadrants, where $\mathrm{{\bf{p}}_{x}\cdot ({\bf{p}}_{y}\times {\bf p}_z)}$ has a + sign.  It follows that $\mathcal{X}$ has a positive value when Ar is driven by synthetic pulse 1. In contrast, in \fig{fig:fig3} for case 2  the probability distribution of the electron momenta p$\mathrm{_{x}}$ and p$\mathrm{_{y}p_{z}}$ has larger values at the second and fourth quadrants, where $\mathrm{{\bf{p}}_{x}\cdot ({\bf{p}}_{y}\times {\bf p}_z)}$ has a - sign. Hence,  $\mathcal{X}$ has a negative value for case 2. The opposite signs of $\mathcal{X}$ when Ar is driven by  electric fields 1 and 2 are consistent with the opposite chirality of these fields. Moreover, $|\mathcal{X}|$ is roughly the same for cases 1 and 2. The small offset is due to the statistical error introduced in our computations from the number of single ionization events considered. This is  also supported by $\mathcal{X}$ being equal to $-0.1\times10^{-2}$, instead of zero, when Ar is driven by the achiral field 3.

Very interestingly, we find that all three measures of chirality $\mathcal{X}\mathrm{(p_{x},p_yp_{z})}$,  $\mathcal{X}\mathrm{(p_{y},p_zp_{x})}$ and  $\mathcal{X}\mathrm{(p_{z},p_x p_{y})}$ have the same values for electric field 
1, 2 and 3. That is, all three measures are equal to $1.8\times10^{-2}$ for electric field 1, equal to -$1.9\times10^{-2}$ for field 2 and equal to -$0.1\times10^{-2}$ for  field 3, see Table \ref{tab:2}. 
The same is true for all three measures of chirality  $\mathcal{X}$ when Ar is driven by synthetic pulses 4,5,6. Indeed, all three $\mathcal{X}$ are equal to 1.$\times10^{-2}$ for electric field 4,
  equal to $-1.1\times10^{-2}$ for electric field 5 and  equal to -$0.1\times10^{-2}$ for electric field 6, see Table II. The above further corroborate that $\mathcal{X}$ is a robust measure of chirality in electron ionization triggered by a chiral field,   yielding
  the same value for any of the three combinations of the components of the final electron momentum.

\begin{table}[ht]
\caption{Measure of chirality $\mathcal{X}$ in electron ionization for globally chiral electric fields 1,2,4,5, and globally achiral fields 3,6.   $\mathcal{X}$ is computed for each of the three $\mathrm{P^m}(\mathrm{p_{x}},\mathrm{p_{y}p_{z}})$, $\mathrm{P^m}(\mathrm{p_{y}},\mathrm{p_{z}p_{x}})$ and $\mathrm{P^m}(\mathrm{p_{z}},\mathrm{p_{x}p_{y}})$, and  m ranges from 1-6. The values are expressed in $10^{-2}.$ } 
\begin{ruledtabular}
\begin{tabular}{lclclclcl}
Case &  $\mathcal{X}\mathrm{\left(p_x, p_yp_z\right)}$ & $\mathcal{X}\mathrm{\left( p_y, p_zp_x\right)}$ & $\mathcal{X}\mathrm{\left( p_z, p_xp_y\right)}$ \\
\hline 
1 & 1.8  &\hspace{0.8cm}  1.8 &\hspace{-0.8cm} 1.8 \\
2 &-1.9 &\hspace{0.8cm}-1.9 &\hspace{-0.8cm}-1.9 \\
3 &-0.1 &\hspace{0.8cm}-0.1 &\hspace{-0.8cm}-0.1 \\
4 & 1.0  &\hspace{0.8cm}  1.0 &\hspace{-0.8cm} 1.0 \\
5 &-1.1 &\hspace{0.8cm}-1.1 &\hspace{-0.8cm}-1.1\\
6 &-0.1 &\hspace{0.8cm}-0.1 &\hspace{-0.8cm}-0.1 \\
\end{tabular}
\end{ruledtabular}
\label{tab:2}
\end{table}

Next, we outline a yet more transparent way to  demonstrate chirality in electron ionization. Namely,  we plot $\mathcal{P}^{\mathrm{m,n}}(\mathrm{p_k,p_ip_j})$ defined as  the difference of the normalized probability distributions in the following way  

\begin{equation}
\mathcal{P}^{\mathrm{m,n}}(\mathrm{p_k,p_ip_j})=\mathrm{P^m(p_{k},\mathrm{p_{i}p_{j}})}-\mathrm{P^n(\mathrm{p_{k}, p_{i}p_{j}})},
\end{equation}
 The corresponding measure of chirality  is given by  
\begin{align}
\begin{split}
&\mathcal{X}d\mathrm{(p_{k},p_{i}p_{j})} = \\
&\mathrm{\iint sign(\mathrm{{\bf{p}}_{k}\cdot ({\bf{p}}_{i}\times {\bf p}_{j})}) \mathcal{P}^\mathrm{m,n}(\mathrm{p_{k}, p_{i}p_{j}}) dp_{i}p_{j}dp_{k} }.\label{eq:mai1d}
\end{split}
\end{align}

\begin{figure} 
\centering
\includegraphics[width=1.\linewidth]{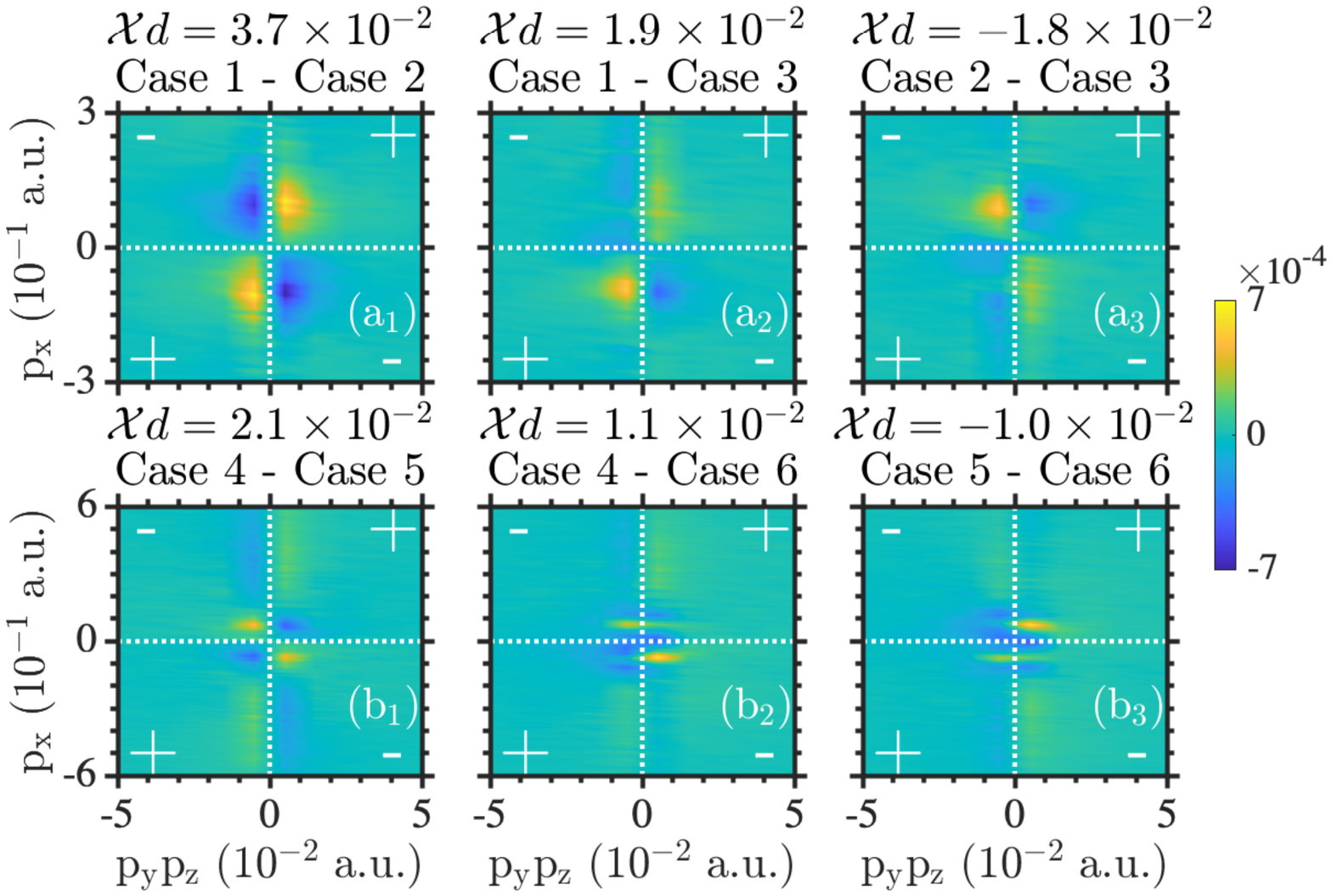}
\caption{Probability distribution $\mathcal{P}^{\mathrm{m,n}}(\mathrm{p_x,p_yp_z})$. The m,n indexes are 1,2 for Case 1-Case 2,
1,3  for Case 1-Case 3 and 2,3 for Case 2-Case 3.  The sign in each quadrant corresponds to the sign of $\mathrm{{\bf{p}}_{x}\cdot ({\bf{p}}_{y}\times {\bf p}_z)}$ in this quadrant. }
\label{fig:fig4}
\end{figure}

 In Figs. \ref{fig:fig4}(a1)-\ref{fig:fig4}(a3), we plot the probability distribution $\mathcal{P}^{\mathrm{m,n}}(\mathrm{p_{x}},\mathrm{p_{y}p_{z}})$ for the pair of opposite chirality electric fields (1,2), i.e. Case1-Case 2 (\fig{fig:fig4}(a1)), and for the pairs of  chiral-achiral  electric fields  (1,3) and (2,3), i.e.  Case 1-Case 3 (\fig{fig:fig4}(a2)) and Case 2-Case 3 ( \fig{fig:fig4}(a3)). 
In each quadrant, we assign the sign resulting from the scalar triple product $\mathrm{{\bf{p}}_{x}\cdot ({\bf{p}}_{y}\times {\bf p}_z)}$. The yellow (blue) color denotes positive (negative) values of $\mathcal{P}^{\mathrm{m,n}}(\mathrm{p_{x}},\mathrm{p_{y}p_{z}})$, corresponding to the electron being more (less) probable to ionize with momenta p$\mathrm{_{x}}$ and p$\mathrm{_{y}p_{z}}$ due to pulse m rather than  pulse n. Next, in each quadrant, we multiply the $\pm$ sign (yellow/blue), resulting from the distribution,    with the $\pm$ sign of $\mathrm{{\bf{p}}_{x}\cdot ({\bf{p}}_{y}\times {\bf p}_z)}$ and then sum up. It easily follows that 
 the measure of chirality  $\mathcal{X}d\mathrm{\left(p_x, p_yp_z\right)}$ is larger and positive ($3.7\times10^{-2}$) for the pair of opposite chirality fields (1,2), see (\fig{fig:fig4}(a1)). Also, $\mathcal{X}d$ is positive ($1.9\times10^{-2}$) for the pair of electric fields (1,3) and negative ($-1.8\times10^{-2}$) for the pair (2,3), with $1.9\times10^{-2}-1.8\times10^{-2}$ being roughly zero, since  pulses 1 and 2 have opposite chirality. As for chirality measures $\mathcal{X}$, we find that all three measures of chirality  $\mathcal{X}d\mathrm{(p_{x},p_yp_{z})}$, $\mathcal{X}d\mathrm{(p_{y},p_zp_{x})}$ and  $\mathcal{X}d\mathrm{(p_{z},p_x p_{y})}$ have the same value for  each of the fields 1,2,3, see Table \ref{tab:3}.
 
\begin{table}[ht]
\caption{Measure of chirality $\mathcal{X}d$   for   globally chiral electric fields 1,2,4,5 and globally achiral fields 3,6.  $\mathcal{X}d$ is computed for each of the three probability distributions 
$\mathcal{P}^{\mathrm{m,n}}(\mathrm{p_{x}},\mathrm{p_{y}p_{z}})$, $\mathcal{P}^{\mathrm{m,n}}(\mathrm{p_{y}},\mathrm{p_{z}p_{x}})$ and $\mathcal{P}^{\mathrm{m,n}}(\mathrm{p_{z}},\mathrm{p_{x}p_{y}})$. The values are expressed in $10^{-2}$. } 
\begin{ruledtabular}
\begin{tabular}{lclclclcl}
m & n &  $\mathcal{X}d\mathrm{\left(p_x, p_yp_z\right)}$ & $\mathcal{X}d\mathrm{\left( p_y, p_zp_x\right)}$ & $\mathcal{X}d\mathrm{\left( p_z, p_xp_y\right)}$ \\
\hline 
1&2 & \hspace{1.2cm} 3.7  & 3.7 &\hspace{0.5cm} 3.7 \\
1&3 & \hspace{1.2cm} 1.9 & 1.9 &\hspace{0.5cm} 1.9 \\
2&3&\hspace{1.2cm}-1.8 &-1.8 &\hspace{0.5cm}-1.8 \\
4&5&\hspace{1.2cm} 2.1 & 2.1 &\hspace{0.5cm} 2.1 \\
4&6&\hspace{1.2cm} 1.1  & 1.1 &\hspace{0.5cm} 1.1 \\
5&6&\hspace{1.2cm}-1.0 &-1.0 &\hspace{0.5cm}-1.0 \\
\end{tabular}
\end{ruledtabular}
\label{tab:3}
\end{table}
A similar  analysis holds for the measures of chirality in electron ionization when Ar  is driven by the globally chiral electric fields 4,5 and the globally achiral field 6. Indeed, in Figs. \ref{fig:fig4}(b1)-\ref{fig:fig4}(b3), we plot  the probability distribution $\mathcal{P}^{\mathrm{m,n}}(\mathrm{p_{x}},\mathrm{p_{y}p_{z}})$ corresponding to the pair of opposite chirality electric fields (4,5), Case 4-Case 5, and to the pairs of chiral-achiral pulses (4,6), Case 4-Case 6, and (5,6), Case 5-Case 6. As for our results for electric fields 1,2,3, we find that $\mathcal{X}d\mathrm{\left(p_x, p_yp_z\right)}$ for the pair of opposite chirality fields (4,5) has the largest value of $2.1\times 10^{-2}$.  
Also, as expected, for the pairs (4,6) and (5,6), we find that $\mathcal{X}d$ has roughly opposite values, $1.1\times 10^{-2}$  and $-1.0\times 10^{-2}$. The exact same results hold for $\mathcal{X}d$ obtained for the other two combinations of  momentum components for fields 4,5,6, see Table \ref{tab:3}.


Summarizing, we  identify a transparent and simple measure of chirality in electron ionization triggered in atoms (Ar) by synthetic pulses. These pulses can create electric fields which are globally chiral or achiral along the focus region. Our computations  account for realistic experimental conditions.  We define this measure  by multiplying 
the sign of the final electron momentum  scalar triple product $\mathrm{{\bf{p}}_{k}\cdot ({\bf{p}}_{i}\times {\bf p}_{j})}$ with the  probability for an electron to ionize with certain values for both p$\mathrm{_{k}}$ and  p$\mathrm{_{i}}$p$\mathrm{_{j}}$. Finally, we integrate over all values of  p$\mathrm{_{k}}$ and p$\mathrm{_{i}}$p$\mathrm{_{j}}$. Three such measures can be defined, corresponding to the three combinations of p$\mathrm{_{k}}$ and p$\mathrm{_{i}}$p$\mathrm{_{j}}$. We find that all three measures of chirality have the same value for a given electric field.   This robust measure of chirality   has opposite values when the electron dynamics is triggered by fields with opposite chirality and is zero for  a globally achiral field.  We expect  that this measure of chirality in electron ionization of atoms  applies to any chiral  field.

\begin{acknowledgments}
The authors A. E and G. P. Katsoulis acknowledge the  use of the UCL Myriad High Throughput Computing Facility (Myriad@UCL), and associated support services, in the completion of this work.
\end{acknowledgments}

\bibliography{bibliography}{}

\end{document}